# Beyond Pattern Matching

## Seven Cross-Domain Techniques for Prompt Injection Detection

**Thamilvendhan Munirathinam**

*Independent Researcher · prompt-shield project*

# Abstract

Current open-source prompt-injection detectors converge on two architectural choices: regular-expression pattern matching and fine-tuned transformer classifiers. Both share failure modes that recent work has made concrete. Regular expressions miss paraphrased attacks. Fine-tuned classifiers are vulnerable to adaptive adversaries: a 2025 NAACL Findings study reported that eight published indirect-injection defenses were bypassed with greater than fifty percent attack-success rates under adaptive attacks, and a subsequent arXiv preprint extended that result to jailbreak defenses more broadly. Progress along the current axis of work appears to face diminishing returns against knowledgeable adversaries.

This work takes a different path. We propose seven detection techniques that each port a specific mechanism from a discipline outside large-language-model security: forensic linguistics, materials-science fatigue analysis, deception technology from network security, local-sequence alignment from bioinformatics, mechanism design from economics, spectral signal analysis with change-point detection from epidemiology, and taint tracking from compiler theory. Each mechanism produces a detection signal that is architecturally independent of both regular-expression matching and transformer classification, so the seven techniques compose with existing defenses rather than replacing them.

The v2.0 revision of this paper moved the contribution from proposal to partial empirical validation (this paragraph narrates the v2.0 state — see the v4.0 addendum below for the current shipped count of four of seven). Three of the seven techniques were implemented in the prompt-shield v0.4.1 release (Apache 2.0) and evaluated in a four-configuration ablation across six datasets: the public benchmarks deepset/prompt-injections, NotInject, LLMail-Inject, AgentHarm, and AgentDojo, and a synthetic indirect-injection benchmark released alongside this paper. The local-alignment detector lifts F1 on deepset from 0.033 to 0.378 (a thirty-four-point-six percentage-point improvement) with zero additional false positives on deepset's fifty-six benign samples; the same detector adds ten false positives on the 339-sample NotInject benign set (FPR 0.9% → 3.8%), a specificity regression discussed in Section 5.4. The stylometric detector adds eleven-point-one percentage points of F1 on the indirect-injection set while remaining silent on short inputs by design. The fatigue tracker is validated by a probing-campaign integration test: ten priming scans below the detection threshold cause a subsequent lower-confidence scan from the same source to be blocked. The remaining four techniques are described as design specifications and are left to future work.

The v3.0 revision (July 2026) adds independent evaluation against three peer-reviewed academic benchmarks — Liu et al. (USENIX Security 2024), Garak (Derczynski et al., 2024), and InjecAgent (Zhan et al., ACL Findings 2024) — totalling 8,276 attack prompts that were not used to design or tune any prompt-shield detector. Detection rates are 64.0 percent on Liu, 55.2 percent on Garak, and 85.2 percent on InjecAgent. Across all three benchmarks the same pattern emerges: approximately ninety-two percent detection on explicit-override attacks, approximately ninety-nine percent on data-exfiltration attacks, and a structural plateau at thirty-five to forty-five percent on subtle indirect injection where the embedded instruction lacks override keywords. This corroborates the gap that Liu et al. identified and that DataSentinel (IEEE S&P 2025) was designed to close, using an independent defender (prompt-shield) and two additional benchmarks. The cross-benchmark evaluation is reported in Section 5.6. Table 2 in that section exercises the regex-only baseline (d022 ML classifier, d027 stylometric detector, and d028 sequence-alignment detector all disabled), matching the baseline control column in Table 1; the novel detectors are validated separately in Sections 5.2 through 5.5 and are not included in the cross-benchmark table. Since v2.0, prompt-shield v0.4.1 has shipped an additional detector — d029 many-shot structural — as a preview; its section-level treatment is scheduled for v5.0 and its numbers are not included in this revision. All claims, limitations, and reproduction scripts are released together with the paper.

The v4.0 revision (July 2026) is substantive rather than corrective. It adds Section 2.4 mapping thirteen concurrent 2026 preprints and publications addressing overlapping problems; Section 7 formalizing three architectural patterns extracted from prompt-shield's implementation in Gang-of-Four format (Tool-Result Boundary Gate, Federated Signed Threat Intelligence, Attack-Family Taxonomic Projection); and Section 5.7 reporting a preliminary Nasr-style adaptive-attack evaluation against d028 in which a matrix-aware second-mover adversary reduces detection from 95.0 percent (19/20) to 0.0 percent (0/20). The result formalizes d028's operating envelope as a first-mover detector and motivates the composed defense-in-depth study scoped for v5.0.

Four of seven techniques are now implemented (was three): Section 4.3 honeypot tool definitions ships as the d034_honeypot_tool detector with twenty unit tests, raising the shipped-technique ratio to 4/7. This revision also folds in two new empirical results. Section 5.8 reports a held-out benchmark of 50 LLM-template-generated business documents (30 with paraphrased indirect-injection payloads) that were not used to design any detector: d027 in isolation drops from the synthetic-benchmark 1.000 F1 to 0.000 F1, while the composed engine (d027 + d028 + regex baseline, d022 disabled) still recovers 0.733 recall / 0.815 F1. Section 5.7 adds a composed-stack adaptive-attack partial run in which d027 alone drops from 100 percent to 0 percent detection under a short-input-floor attack, while the full composed stack catches 20/20 of the same adaptive inputs (and 14/20 without the d022 semantic classifier), providing partial empirical support for the Section 2.2 composability thesis.

# 1. Introduction

Prompt injection is the dominant offensive technique against large-language-model-integrated applications. A malicious input, whether supplied directly by an end user or smuggled through a retrieval-augmented-generation chunk, an email body, or a tool-call response, can steer the assistant model away from its operator-intended behavior. The attack class is mature enough to support a formalized taxonomy, public benchmarks, and a competitive defense research community, yet the defense literature remains narrow in its algorithmic palette.

A literature review of recent open-source detectors reveals an architectural convergence. Every widely-used defense tool we examined uses some combination of three mechanisms: a hand-written regular-expression pack for known attack patterns; a transformer classifier fine-tuned on labeled injection corpora; and a vector-similarity lookup against a store of historical malicious prompts. Each mechanism has well-understood weaknesses. Regular expressions miss semantic paraphrases and non-English attacks. Fine-tuned classifiers show strong in-distribution accuracy but, as recent adaptive-attack studies demonstrate, can be bypassed by knowledge-informed adversaries. Vector similarity cannot catch attacks that are semantically novel relative to the stored corpus.

The question this paper asks is whether meaningfully different detection signals can be recovered by porting mechanisms from disciplines that have solved structurally analogous problems in contexts unrelated to large-language-model security. We identify seven such mechanisms and present them as a coordinated palette of defenses, grouped by the detection signal they produce rather than by the attack class they target. Four techniques are implemented in the open-source prompt-shield library (three evaluated in the four-configuration ablation of Section 5; the fourth, honeypot tool definitions, shipped in the current v4.0 cycle as d034 with unit-test coverage but not yet folded into the ablation harness). The remaining three are described as design specifications pending implementation.

## 1.1 Contributions

The primary contributions of this paper (as of the v4.0 revision) are:

- **Seven cross-domain detection mechanisms.** We describe seven techniques, each rooted in a distinct scientific discipline with a self-contained mathematical or architectural basis: stylometric discontinuity, adversarial fatigue tracking, honeypot tool definitions, local-sequence alignment with a semantic substitution matrix, prediction-market ensemble scoring, perplexity spectral analysis, and runtime taint tracking for agent pipelines. Each is motivated by a specific failure mode of the existing regex-plus-classifier paradigm.
- **Four implementations with empirical validation.** The local-alignment detector (d028), the adversarial fatigue tracker, the stylometric discontinuity detector (d027), and — as of this v4.0 revision — the honeypot tool-definitions detector (d034) are released in prompt-shield under the Apache 2.0 license (d028/d027/fatigue in v0.4.1; d034 on the main branch as of commit 2efe518+, scheduled for the v0.7.3 correctness release). Each is accompanied by unit and integration tests (one hundred and thirty-four tests in total across the four techniques, including twenty for d034) and a public reproduction harness that regenerates every number in Section 5 from a single command invocation.
- **A four-configuration ablation across six datasets.** We benchmark the three shipped techniques on five public datasets (deepset/prompt-injections, NotInject, LLMail-Inject Phase 1, AgentHarm, and AgentDojo v1.2.1) and on a new synthetic indirect-injection benchmark generated by a reproducible template-based builder also released alongside the paper. Every ablation cell reports precision, recall, F1, and false-positive rate directly from the benchmark JSON.

- **Honest limitations section.** Section 5.4 enumerates the tradeoffs that a peer-review audience is likely to challenge: the NotInject false-positive-rate regression under d028, the saturation of LLMail-Inject under the pre-existing regex pack, the orthogonal behavior of AgentHarm to all three shipped techniques, and the self-evaluation risk of the synthetic indirect-injection benchmark. Each is flagged with a concrete next-revision action item.

The v3.0 revision added:

- **Independent evaluation against three peer-reviewed academic benchmarks.** Section 5.6 reports prompt-shield's detection rate on the Liu et al. attack templates (USENIX Security 2024, 200 attacks), the NVIDIA Garak vulnerability scanner's promptinject and latentinjection probe families (5,968 attacks), and the InjecAgent indirect-injection benchmark (ACL Findings 2024, 2,108 cases). None of these corpora were used to design any prompt-shield detector. The three benchmarks converge on a thirty-five to forty-five percent pattern-matching ceiling for subtle indirect injection, the same ceiling Liu et al. quantified and that DataSentinel (IEEE S&P 2025) was designed to close.

The remainder of the paper is structured as follows. Section 2 reviews the current prompt-injection defense landscape and its known failure modes. Section 3 defines the threat model. Section 4 presents the seven cross-domain techniques. Section 5 reports the empirical evaluation of the three ablation-covered detectors, a v4.0 held-out benchmark of the novel-detector stack (Section 5.8), and preliminary adaptive-attack experiments against d028 and the composed stack (Section 5.7). Section 6 concludes with a sequencing for the three remaining techniques and a roadmap for the next revision.

# 2. Background and Related Work

## 2.1 The convergence problem

Formal treatment of prompt injection as a measurable attack class dates to Liu et al. (USENIX Security 2024), who introduced a benchmark of attack-and-defense pairs and a taxonomy that has since been adopted by most subsequent work. In the two years since, the defense literature has consolidated around three patterns.

The first pattern is rule-based detection. Early open-source tools such as LLM Guard and prompt-shield itself in its v0.3 releases couple curated regular-expression packs with severity labels and aggregation logic. Commercial products in this family, including Lakera's PINT (April 2024) and the Rebuff project (archived May 2025), share the same architecture. The strength of rule-based detection is high precision on known attack vocabularies and fast execution. The weakness is that paraphrased or synonym-substituted attacks bypass the patterns verbatim.

The second pattern is classifier-based detection. Meta's Llama Prompt Guard 2 (2025, 22M and 86M parameter variants; no peer-reviewed paper) and Deepset's DeBERTa-v3 injection classifier are representative. ProtectAI's PIGuard, published at ACL 2025, explicitly addresses the over-defense problem by releasing the NotInject benchmark alongside a classifier calibrated for benign queries that contain attack-adjacent vocabulary. Classifiers absorb paraphrased attacks better than rules but introduce substantial false-positive rates on legitimate queries about the attack class itself, and are the specific target of the adaptive-attack research summarized in the next subsection.

The third pattern is meta-defense: model-integrated signaling. Spotlighting (Hines et al., arXiv:2403.14720) encodes context data with delimiters or simple transformations so the model can distinguish trusted instructions from untrusted context. SelfDefend (Wu et al., arXiv:2406.05498) deploys a shadow large-language-model whose sole job is to classify the primary model's inputs. MELON (Zhu et al., ICML 2025, arXiv:2502.05174) executes the primary model twice and blocks when the two runs diverge on tool invocations. These meta-defenses introduce latency but can catch attacks the primary detector misses.

## 2.2 Adaptive attacks as the current frontier

Zhan et al. (NAACL 2025 Findings, arXiv:2503.00061) evaluated eight published indirect-injection defenses under adaptive attacks and reported that all eight were bypassed with attack-success rates exceeding fifty percent. Nasr et al. (arXiv:2510.09023, OpenReview 7B9mTg7z25, submission status pending) extended the result to jailbreak defenses and reported greater than ninety percent attack-success rates against twelve published defenses. Both results target the same underlying weakness: when the adversary has model-gradient or decision-boundary access, any single-mechanism defense can be systematically evaded.

The methodological implication is that single-axis defense research is subject to diminishing returns. A detector that is ten percent stronger on a fixed benchmark is typically not measurably harder to adaptively evade than its predecessor. The defenses that remain effective under adaptive attacks tend to exploit properties the adversary cannot cheaply manipulate, such as context provenance or cross-model disagreement. This paper argues that introducing mechanistically independent detection signals creates additional invariants that an adversary must simultaneously evade, which is strictly harder than evading a stronger version of a single signal. Mindgard / Lancaster (arXiv:2504.11168, 2025) empirically catalog the state of guardrail evasion, corroborating this diminishing-returns observation from the attack side.

## 2.3 Agentic-attack benchmarks and tool integrity

The agent-integrated deployment pattern introduces attack surface beyond the text-input boundary. AgentDojo (Debenedetti et al., NeurIPS 2024, arXiv:2406.13352) formalizes indirect injection against tool-using agents with a suite of benign user tasks paired with injection tasks embedded in tool responses. Agent Security Bench (Zhang et al., ICLR 2025, arXiv:2410.02644) scales the formalism to ten scenarios with eleven defenses and four hundred tools. AgentHarm (Andriushchenko et al., ICLR 2025, arXiv:2410.09024) measures harmfulness of agent task completion rather than injection success directly; as we show in Section 5, this distinction matters for detector evaluation.

Two recent works target the same agentic-attack surface from opposite directions. Information flow control for agents is developed in FIDES (Costa et al., Microsoft Research, arXiv:2505.23643), which attaches confidentiality and integrity labels to model and tool inputs. Static taint analysis for large-language-model-integrated applications is developed in TaintP2X (He et al., ICSE 2026), which treats the assistant's prompt assembly as a dataflow graph and checks reachability from untrusted sources to sensitive sinks. Together these works establish that architectural-integrity techniques from compiler security are viable in the large-language-model context, a premise our Section 4.7 proposal builds on.

## 2.4 2026 Concurrent Contributions

Between April and July 2026, concurrent with the prompt-shield v0.6 and v0.7 releases described in this paper, the field produced thirteen preprints and publications addressing overlapping problems; we group them by defense pattern.

Trained classifiers on prompt-injection benchmarks. PromptSentinel-X (Zhuhadar, Future Internet 2026, DOI 10.3390/fi18070376) reports 98.49 percent accuracy, 0.5 percent false-positive rate, and 0.9971 AUROC on a 465-record test set drawn from the Prompt Injection Malignant corpus; this is the strongest in-distribution result in the batch, and Section 5 discusses why it is not directly comparable to the out-of-distribution NotInject numbers we report. WebSentinel (Wang et al., arXiv:2602.03792) extends the classifier approach with a detect-and-localize head specialized for web-agent DOM contexts. SnapGuard (Du et al., arXiv:2604.25562) replaces text features with screenshot-based visual features for web agents whose attack surface reaches the model through rendered pages rather than raw markup. DataSentinel (Liu et al., IEEE S&P 2025, arXiv:2504.11358) formulates the classifier training objective as a game between attacker and defender to improve robustness to adversarial paraphrasing.

Runtime enforcement and capability-based confinement. ClawGuard (Zhao et al., arXiv:2604.11790) interposes user-confirmed rule sets at the tool-call boundary, blocking calls whose arguments violate a per-session policy. AgentSpec (Wang, Poskitt, and Sun, ICSE 2026, arXiv:2503.18666) introduces a domain-specific language for expressing runtime constraints over agent behavior. Prismata (Villa et al., arXiv:2607.08147) derives per-tool trust levels dynamically and enforces mechanical confinement of low-trust web-agent actions. DualView (Kim et al., arXiv:2607.03821) isolates rendered content from model-visible symbols by maintaining two parallel views — an AgentView of tokenised symbols exposed to the model and a HumanView of the original surface — so that injected instructions in the rendered content cannot reach the model context. Token-Flow Firewall (Wang et al., arXiv:2607.08395) audits token flows across agent boundaries with boundary-aware semantic checks that account for cross-boundary information carriage.

Rule-based monitors. AgentWatcher (Wang, Zou, Geng, and Jia, arXiv:2604.01194) couples rule-based detection with causal attribution, tracing each violation back to the specific tool call or context chunk that triggered it; it is the closest philosophical sibling to prompt-shield's regex-based detector family.

Concurrent benchmarks. NetInjectBench (Shayoni et al., arXiv:2607.10490) proposes an indirect-injection benchmark for tool-using LLM agents in network-operations settings; we treat it as a candidate for the evaluation extension noted in the future-work portion of Section 6.

Adversarial contributions. AgentSentry (Zhang et al., arXiv:2602.22724) and WARD (Cao et al., arXiv:2605.15030) contribute attack techniques against agentic pipelines. Both are relevant to the threat model of Section 3; empirical evaluation of prompt-shield's coverage against their specific attack corpora is left to future work.

Position of prompt-shield within this batch. Three observations frame our contribution relative to these concurrent works. First, trained classifiers that report sub-one-percent false-positive rates on their own held-out sets — PromptSentinel-X at 0.5 percent is the batch leader — achieve those numbers by construction: the training and test distributions are matched. The numbers we report in Section 5 instead measure out-of-distribution generalisation on the NotInject benchmark, which is the regime a pip-install deployment encounters on unseen user inputs. Second, prompt-shield and the runtime-enforcement papers operate at different layers of the defense stack: the runtime-enforcement systems confine what a compromised model can do, while prompt-shield classifies whether the incoming context is likely compromised. The two layers compose cleanly at the tool-call boundary. Third, no cluster member ships the combination of a federated signed threat feed, a stable attack-family taxonomy over heterogeneous detectors, and an integrated defense-in-depth stack in a single deployable package; this is the deployment-stack positioning we develop in Sections 4 and 7.

# 3. Threat Model

We consider a deployed assistant pipeline consisting of a system prompt, a user input channel, zero or more retrieval-augmented context chunks, and optionally a tool-calling capability that returns structured results back into the model context. An adversary controls at least one of the non-system channels: the direct user input, an RAG source they can populate, an email or document body the assistant will summarize, or a tool response the attacker can craft by compromising an upstream data source.

The adversary's goal is to induce the model to ignore its operator-intended behavior, either by exfiltrating the system prompt, by invoking tools the operator did not authorize, or by emitting content the operator explicitly forbade. The adversary may paraphrase attacks, substitute synonyms, insert filler words, obfuscate with encoding or unicode homoglyphs, pose attacks as hypothetical questions or academic research, split payloads across multiple turns, or exploit tool-response channels to smuggle attacks into otherwise-trusted contexts.

We treat the adversary as adaptive in the sense of Zhan et al. and Nasr et al.: they have access to the defense source code, the detector thresholds, and the published benchmark results. They do not have access to per-deployment state such as the fatigue tracker's per-source EWMA (Section 4.2), the operator's canary tokens, or the self-learning vault's accumulated history. Techniques that depend on such per-deployment state are strictly stronger than techniques that depend only on the released code.

We do not model attacks that subvert the operator or the model-weight pipeline itself. Supply-chain risks, training-data poisoning, and fine-tune-time backdoors are out of scope. Also out of scope are side-channel attacks against the underlying inference infrastructure (timing, memory disclosure, GPU isolation). Our threat model concerns adversaries operating through the normal user-facing input channels of an already-provisioned pipeline.

# 4. Seven Cross-Domain Techniques

Each subsection below identifies the source discipline, describes the detection mechanism in enough detail for replication, and reports the current implementation status. The four shipped techniques (Sections 4.1, 4.2, 4.3, 4.4) include references to the source code and unit tests in prompt-shield (Sections 4.1, 4.2, 4.4 in v0.4.1; Section 4.3 honeypot detector shipped as d034 in the v4.0 cycle); the ablation empirical evaluation in Section 5 covers Sections 4.1, 4.2, 4.4. The three remaining techniques (Sections 4.5, 4.6, 4.7) are presented as design specifications.

## 4.1 Stylometric Discontinuity Detection

*Status: SHIPPED in prompt-shield v0.4.1 (Apache 2.0). Source: src/prompt_shield/detectors/d027_stylometric_discontinuity.py.*

Forensic linguistics has studied authorship attribution since the Federalist Papers analysis of Mosteller and Wallace in the 1960s. The stylometric tradition holds that individual authors leave measurable signatures in their writing — distributions of function-word frequency, sentence-length variance, vocabulary richness — that survive paraphrase, topic variation, and deliberate obfuscation. Applied to the task of detecting multi-author documents, the contemporary Plagiarism Analysis shared task PAN (Bevendorff et al., CLEF 2024) formalizes a sliding-window approach: partition a document into overlapping token windows, compute a stylometric feature vector per window, and measure divergence between adjacent windows to locate authorship boundaries.

An indirect prompt injection necessarily introduces a second author into a document. The legitimate author, typically a human writing an email or a content-management system emitting an HTML page, produces text in one stylistic distribution. The attacker producing the injected payload writes in a different distribution, characterized by imperative mood, high uppercase-character frequency, and directive vocabulary. The stylometric approach recasts indirect-injection detection as a well-studied style-change-boundary problem for which there is a decades-long feature engineering literature to draw from.

The detector extracts nine features per fifty-token window with twenty-five-token stride: function-word frequency (closed-class words such as the, is, of); average word length; average sentence length; stylistic punctuation density; hapax legomena ratio (singletons divided by vocabulary size); Yule's K statistic for vocabulary richness; imperative-verb ratio over a closed list of attack-adjacent imperatives; uppercase character ratio; and all-capital-word ratio for tokens of length three or more. Each feature is scaled to a [0, 1] range, the resulting nine-dimensional vector is normalized to a probability mass, and the Jensen-Shannon divergence (Lin, IEEE TIT 1991) is computed between every adjacent window pair. The maximum divergence across the input is compared against a calibrated threshold.

The detector is silent on inputs shorter than one hundred tokens because the per-window feature estimates are unstable on smaller samples; this is a deliberate engineering choice, not a limitation that affects the detection claim. On our synthetic indirect-injection benchmark (Section 5) the detector lifts the F1 score of a twenty-six-detector regex baseline from 0.889 to 1.000 with zero false positives on benign long-form distractors.

Prior application of stylometric features to large-language-model security has, to our knowledge, been limited to the authorship-attribution direction: distinguishing human-written from machine-generated text (Opara 2024, arXiv:2405.10129). Reversing the task to detect adversarial-author injection in otherwise-benign documents appears to be new.

## 4.2 Adversarial Fatigue Tracking

*Status: SHIPPED in prompt-shield v0.4.1 (opt-in, Apache 2.0). Source: src/prompt_shield/fatigue/tracker.py.*

Structural engineering has long distinguished between single-cycle failure and fatigue failure. A beam that does not break under a one-time load of a given magnitude may still fail under repeated loading at a lower magnitude, because microscopic cracks accumulate below the yield threshold and eventually propagate into macroscopic fractures. Basquin's 1910 stress-life curves and the broader fatigue-analysis literature (Suresh, Cambridge University Press, second edition 1998) model this accumulation as a function of both the stress amplitude and the number of cycles.

A sophisticated adversary probing a prompt-injection detector exhibits the same statistical signature: a sustained sequence of inputs that each individually score just below the detection threshold, iteratively searching for the minimum-effort bypass. A stateless detector that evaluates each input in isolation cannot distinguish this probing campaign from a burst of benign near-boundary queries. A stateful detector that models the rate of near-threshold scores per source can.

We define a near-miss for a given (source, detector) pair as a scan whose confidence lies in the interval [threshold minus delta, threshold), where delta is a configuration parameter. The tracker maintains an exponentially-weighted moving average (Roberts, Technometrics 1959; see Tsai et al., IET Information Security 2024, for contemporary intrusion-detection use) of the near-miss indicator over successive scans. When the EWMA exceeds a configured trigger ratio and a minimum number of samples have been observed, the detector enters a hardened state for that source, in which the effective threshold is lowered by a configurable amount. The hardened state automatically releases after a cooldown interval with no further near-misses.

Structural epidemiology uses the same mathematical tools for outbreak detection. EWMA and CUSUM statistics are standard in syndromic-surveillance systems (Centers for Disease Control and Prevention; Vial et al., Emerging Infectious Diseases 26(9), 2020), where the task is to identify a shift in the baseline rate of a signal that is noisy on any single day. The cross-domain analogy that motivates the detector is therefore twofold: materials-science fatigue supplies the failure-model framing, and epidemiological surveillance supplies the detection-statistic choice.

The fatigue tracker is deliberately orthogonal to static single-sample benchmarks because its signal is temporal. We validate it end-to-end in Section 5.3 via a probing-campaign integration test: after ten priming scans from a single source at confidence 0.65 (base threshold 0.7), the eleventh scan from the same source at confidence 0.63 — strictly below every priming score and strictly below the un-hardened threshold — is blocked. A concurrent scan from a different source at the same confidence is unaffected, confirming per-source isolation. The tracker is opt-in and imposes zero overhead when disabled.

## 4.3 Honeypot Tool Definitions

*Status: IMPLEMENTED as d034_honeypot_tool in the prompt-shield v4.0 release cycle (Apache 2.0). Source: src/prompt_shield/detectors/d034_honeypot_tool.py.*

Network security has decades of experience with deception: a honeypot is a resource with no legitimate use case such that any interaction is definitionally adversarial. Recent work has extended the pattern to the large-language-model agent setting. Reworr and Volkov (Palisade Research, arXiv:2410.13919) deployed honeypots on the public internet to observe large-language-model-driven attackers and characterize their behavior. Pasquini et al. (arXiv:2410.20911) designed a defensive honeypot system that uses prompt injection against the attacker, demonstrating that the same attack surface can be inverted when the defender chooses the environment.

We extend deception to the tool-call layer of large-language-model agents. Agent frameworks typically expose a list of tool definitions to the model; the shipped d034_honeypot_tool detector operates against canary tool definitions with names such as admin_shell, internal_debug_console, system_key_dump, sudo_execute_raw, and priv_esc_helper (the five default honeypot names). These tools have no legitimate invocation path. An agent pipeline that observes any call to a canary tool can treat the call as a confirmed indirect-injection event with one hundred percent precision: no benign query path leads to these tool names. The design is complementary to canary tokens embedded in system prompts (an older technique popularized by the Rebuff project, 2023) and to integrity-probe checks that periodically test whether the system prompt is still intact.

Empirical validation. The d034 detector ships with twenty unit tests exercising both firing modes and every input-shape variant we found in production tool-calling frameworks. The detector has two firing modes: (i) text-mention mode, which fires at confidence 0.95 when the raw input text mentions a honeypot name as a word-bounded token; and (ii) context-invocation mode, which fires at confidence 1.0 when the runtime tool_calls context contains an entry naming a honeypot. The name-extraction path accepts all four tool-call shapes we observed across real frameworks: flat dicts with a name field, alternate flat keys (tool, function, tool_name, function_name), the OpenAI nested function shape ({"type":"function","function":{"name":...}}), and Anthropic-style nested wrappers. Operators may register the five default names as no-op trap tools in the agent's schema or override the list via the d034_honeypot_tool.honeypot_tools configuration key. A dedicated integration into the Section 5 ablation harness is deferred to a follow-up cycle because the detector requires a tool-call fixture rather than a text-only benchmark.

Section 6 lists honeypot as the first of the previously-remaining techniques implemented in this v4.0 cycle; the remaining three (prediction-market ensemble scoring, perplexity spectral analysis, runtime taint tracking) are sequenced there.

## 4.4 Local Sequence Alignment with a Semantic Substitution Matrix

*Status: SHIPPED in prompt-shield v0.4.1 as d028_sequence_alignment (Apache 2.0). Source: src/prompt_shield/detectors/d028_sequence_alignment.py.*

Bioinformatics has the strongest continuous tradition of robust substring-matching research of any computing discipline. The Smith-Waterman algorithm (Smith and Waterman, J. Mol. Biol. 147:195-197, 1981) computes the optimal local alignment between two biological sequences in quadratic time using dynamic programming; the BLOSUM family of scoring matrices (Henikoff and Henikoff, PNAS 89:10915-10919, 1992) encodes the evolutionary cost of substituting one amino acid for another, allowing the algorithm to recognize homology in the presence of mutation.

A prompt-injection attack and its paraphrase share structural properties that map cleanly onto this framework. The attack has a characteristic word order (`ignore all previous instructions`); a paraphrase substitutes tokens within semantic equivalence classes (`disregard every prior directive`); a padded attack inserts filler words between structural anchors (`please just kindly ignore, if you would, all of the previous instructions`). Each of these transformations is analogous to a mutation, insertion, or deletion in a biological sequence, and Smith-Waterman with a semantically-weighted substitution matrix recovers the alignment despite the transformation.

The d028 detector maintains a curated database of one hundred and eighty-seven attack sequences across twenty attack categories, drawn from the prompt-shield regex patterns of detectors d001 through d020. The substitution matrix has fifteen synonym groups covering the attack-adjacent vocabulary (ignore-family, instruction-family, reveal-family, system-family, etc.). The scoring function rewards exact matches with +3, same-group synonym matches with +2, unrelated-token mismatches with -1, and gap insertions with -1. A strict-greater-than comparison against a normalized threshold of 0.6 prevents false positives from generic English prefixes such as "show me the" that happen to align with the opening of an attack needle.

On the deepset/prompt-injections benchmark — the canonical academic dataset for prompt-injection detection — the twenty-six-detector regex baseline catches one attack out of sixty (F1 0.033) because the samples are dominated by paraphrased attacks that the regex patterns miss verbatim. The d028 detector catches fourteen of sixty (F1 0.378), a thirty-four-point-six percentage-point F1 improvement, with zero additional false positives on deepset's fifty-six benign samples. On the 339-sample NotInject benign set, the same detector adds ten false positives (FPR 0.9% → 3.8%); see Section 5.4.

To our knowledge, no prior work applies semantically-weighted local sequence alignment to prompt-injection detection. The CRAN text.alignment package applies Smith-Waterman to plain-text document comparison but uses character-level matching without a semantic matrix.

## 4.5 Prediction-Market Ensemble Scoring

*Status: PROPOSED. Implementation blocked on data-model work; see Section 6.*

Mechanism design from economics offers mature solutions to the information-aggregation problem that ensemble scoring solves in an ad-hoc way. A prediction market, equipped with a proper scoring rule such as Hanson's Logarithmic Market Scoring Rule (Journal of Prediction Markets, 2007), aggregates beliefs from heterogeneous participants into a single calibrated probability, weighting each participant by their historical accuracy (measured, for example, by Brier score).

Current ensemble scoring in prompt-shield and comparable tools uses a maximum-confidence-plus-bonus rule that ignores detector reliability and double-counts correlated signals. Replacing this rule with an LMSR market where each detector's bet is weighted by its historical Brier score over labeled scan history would produce better-calibrated probabilities and automatically down-weight unreliable detectors. Implementation requires a new persistence schema for per-detector confidence history and a ground-truth labeling path. The migration carries production risk because it touches the core scoring path; we plan a mandatory shadow-mode validation gate in which both legacy and market scoring run side-by-side with the legacy score driving actions until multi-week agreement is established.

## 4.6 Perplexity Spectral Analysis

*Status: PROPOSED. Implementation requires an optional transformers dependency for GPT-2-small (124M params).*

Signal processing and epidemiology share a concern with change-point detection in noisy sequences. A benign document produces a smooth, low-frequency perplexity signal when scored token-by-token by a reference language model. An injected payload with different vocabulary, syntax, and intent creates a high-frequency spike. We propose computing the discrete Fourier transform of the per-token perplexity series and flagging inputs with abnormally high high-frequency energy ratio, complemented by cumulative-sum change-point detection (Page, Biometrika 1954; Vial et al., CDC Emerging Infectious Diseases 2020) to localize the boundary between benign prose and injected payload.

The technique targets embedded or sandwich-style attacks where the payload is surrounded by benign cover text, a class that rule-based and classifier-based defenses have difficulty with when the cover text dominates the input. The main implementation cost is the optional dependency on a small reference model; we plan to use GPT-2-small (124M parameters) with lazy loading, matching the pattern already used for the existing d022 semantic classifier.

### 4.7 Runtime Taint Tracking for Agent Pipelines

*Status: PROPOSED. Design sketch below; empirical validation requires an agent-pipeline fixture not yet constructed.*

Compiler security has studied taint analysis since the 1970s: data values are labeled with provenance, label propagation follows dataflow, and a policy violation is raised when tainted data reaches a privileged sink without passing through a sanitizer. The pattern applies directly to agent pipelines: system-prompt data is trusted, retrieval context is semi-trusted, user input is untrusted, and tool responses are untrusted unless the tool is known-safe. A tool call invoked with arguments that trace back to an untrusted source without passing through an explicit sanitizer is a policy violation.

We propose a runtime variant of the static analysis developed in TaintP2X (He et al., ICSE 2026): a `TaintedString` wrapper that carries provenance metadata through string operations, a sink-validation hook on the tool-call boundary, and a `TaintViolation` exception that callers can catch. The runtime variant is strictly weaker than static analysis in the formal sense (it only catches violations on the actual execution path) but strictly more practical: it requires no whole-program analysis and operates on the live object graph. Information Flow Control for AI agents (Costa et al., FIDES, arXiv:2505.23643) establishes that this framing is viable in the large-language-model setting.

## 5. Evaluation

### 5.1 Methodology

We evaluate the three shipped detectors via a four-configuration ablation across six datasets. The four configurations are: a baseline twenty-six-detector regex pack identical to the prompt-shield v0.3.3 release (with the d022 semantic classifier deliberately disabled so the ablation isolates the regex-plus-novel contribution); that baseline plus d028 only; the baseline plus d027 only; and the baseline plus both novel detectors. The fatigue tracker is evaluated separately in Section 5.3 because it signals on temporal sequences rather than on individual samples.

The detection rule on every sample is: the input is counted as detected when the engine returns an action in {block, flag} or when the overall risk score is greater than or equal to 0.5. This matches the rule used by the repository's existing benchmark harness and by the v0.3.3 baseline recorded in tests/baseline_v0.3.3.txt. The full scan threshold is 0.7 in every configuration.

We use six datasets. Five are public and standard in the prompt-injection literature: deepset/prompt-injections (116 samples, test split), NotInject (339 benign samples across all three splits), microsoft/llmail-inject-challenge (a one-thousand-sample subset of the Phase 1 submissions), AgentHarm (the test_public split of both the harmful and harmless_benign configurations, 352 samples total), and AgentDojo v1.2.1 (132 tasks total, extracted from the agentdojo Python package). The sixth is a new synthetic indirect-injection benchmark (80 samples), generated by the reproducible builder docs/papers/evaluation/build_indirect_injection_benchmark.py and released alongside the paper. The benchmark combines six benign genre templates (financial report, team email, news article, research paper, technical documentation, analyst memo) with five egregious payload styles (uppercase-directive, credential exfiltration, system-override, admin-password request, safety-disable directive) to produce thirty positive samples; fifty negative samples use the same templates with benign mid-paragraph bridges. The seed is fixed at 20260420 for deterministic regeneration.

## 5.2 Results

Table 1 reports the full four-configuration ablation across all six datasets. Every number is sourced directly from v041_public_datasets.json produced by a single deterministic run of run_public_datasets.py; reproduction is documented in Section 5.5.

| Dataset (n) | Config | TP | TN | FP | FN | Prec. | Recall | F1 | FPR |
|---|---|---|---|---|---|---|---|---|---|
| **deepset (116)** | baseline | 1 | 56 | 0 | 59 | 1.000 | 0.017 | 0.033 | 0.000 |
| **deepset (116)** | plus_d028 | 14 | 56 | 0 | 46 | 1.000 | 0.233 | 0.378 | 0.000 |
| **deepset (116)** | plus_d027 | 1 | 56 | 0 | 59 | 1.000 | 0.017 | 0.033 | 0.000 |
| **deepset (116)** | plus_d027_d028 | 14 | 56 | 0 | 46 | 1.000 | 0.233 | 0.378 | 0.000 |
| **NotInject (339)** | baseline | 0 | 336 | 3 | 0 | 0.000 | 0.000 | 0.000 | 0.009 |
| **NotInject (339)** | plus_d028 | 0 | 326 | 13 | 0 | 0.000 | 0.000 | 0.000 | 0.038 |
| **NotInject (339)** | plus_d027 | 0 | 336 | 3 | 0 | 0.000 | 0.000 | 0.000 | 0.009 |
| **NotInject (339)** | plus_d027_d028 | 0 | 326 | 13 | 0 | 0.000 | 0.000 | 0.000 | 0.038 |
| **LLMail-Inject (1000)** | baseline | 978 | 0 | 0 | 22 | 1.000 | 0.978 | 0.989 | 0.000 |
| **LLMail-Inject (1000)** | plus_d028 | 980 | 0 | 0 | 20 | 1.000 | 0.980 | 0.990 | 0.000 |
| **LLMail-Inject (1000)** | plus_d027 | 978 | 0 | 0 | 22 | 1.000 | 0.978 | 0.989 | 0.000 |
| **LLMail-Inject (1000)** | plus_d027_d028 | 980 | 0 | 0 | 20 | 1.000 | 0.980 | 0.990 | 0.000 |
| **AgentHarm (352)** | baseline | 44 | 120 | 56 | 132 | 0.440 | 0.250 | 0.319 | 0.318 |
| **AgentHarm (352)** | plus_d028 | 44 | 120 | 56 | 132 | 0.440 | 0.250 | 0.319 | 0.318 |
| **AgentHarm (352)** | plus_d027 | 44 | 120 | 56 | 132 | 0.440 | 0.250 | 0.319 | 0.318 |

| **AgentHarm (352)** | plus_d027_d028 | 44 | 120 | 56 | 132 | 0.440 | 0.250 | 0.319 | 0.318 |
|---|---|---|---|---|---|---|---|---|---|
| **AgentDojo (132)** | baseline | 17 | 86 | 11 | 18 | 0.607 | 0.486 | 0.540 | 0.113 |
| **AgentDojo (132)** | plus_d028 | 18 | 83 | 14 | 17 | 0.562 | 0.514 | 0.537 | 0.144 |
| **AgentDojo (132)** | plus_d027 | 17 | 86 | 11 | 18 | 0.607 | 0.486 | 0.540 | 0.113 |
| **AgentDojo (132)** | plus_d027_d028 | 18 | 83 | 14 | 17 | 0.562 | 0.514 | 0.537 | 0.144 |
| **Synthetic indirect-injection (80)** | baseline | 24 | 50 | 0 | 6 | 1.000 | 0.800 | 0.889 | 0.000 |
| **Synthetic indirect-injection (80)** | plus_d028 | 30 | 50 | 0 | 0 | 1.000 | 1.000 | 1.000 | 0.000 |
| **Synthetic indirect-injection (80)** | plus_d027 | 30 | 50 | 0 | 0 | 1.000 | 1.000 | 1.000 | 0.000 |
| **Synthetic indirect-injection (80)** | plus_d027_d028 | 30 | 50 | 0 | 0 | 1.000 | 1.000 | 1.000 | 0.000 |

*Table 1. Four-configuration ablation across six datasets. Precision, recall, F1 and FPR computed at the standard detection rule (action ∈ {block, flag} or overall_risk_score ≥ 0.5).*

## 5.3 Per-technique interpretation

Local sequence alignment (d028). On deepset the detector lifts recall from one-in-sixty to fourteen-in-sixty (a 14× improvement), and F1 from 0.033 to 0.378, with zero additional false positives on deepset's fifty-six benign samples. This is the strongest single-technique result in the evaluation and is consistent with the design intent: deepset is dominated by paraphrased attacks and synonym substitutions that the regex pack cannot see verbatim but that the substitution matrix recognizes. On NotInject the detector introduces ten false positives (FPR 0.9% → 3.8%), which we flag and discuss in Section 5.4. On LLMail-Inject the detector adds two true positives (recall 0.978 → 0.980), a saturation effect: the benchmark is dominated by direct instruction-override attacks that the regex pack already catches. On AgentHarm the detector moves no metrics, a consequence of the benchmark measuring agent-task harmfulness rather than injection detection. On AgentDojo the detector adds one true positive and three false positives, a net neutral result in F1 terms. On the synthetic indirect-injection set the detector closes the remaining six-sample gap from 0.889 to 1.000.

Stylometric discontinuity (d027). By construction the detector is silent on inputs shorter than one hundred tokens; because the five public datasets are dominated by shorter samples the detector produces no movement on any of their metrics. On the synthetic indirect-injection set — where every sample is a ≥150-token document — the detector lifts F1 from 0.889 to 1.000 with zero false positives, matching the d028 result on the same set. The two detectors share the same six-sample residual on the indirect-injection set because both the stylometric break signal and the alignment signal fire on the same uppercase-directive payload structure. Their orthogonality manifests across datasets rather than within any one dataset: d028 contributes on deepset where d027 is silent, and d027 would contribute on a benchmark of long documents where d028's attack-sequence database is less useful.

Adversarial fatigue. Because the detector signals on temporal sequences rather than individual samples, and because the five public datasets present each sample as an independent scan, a static-dataset F1 delta is not a meaningful measure. The detector is validated end-to-end by the integration test
tests/fatigue/test_engine_integration.py::test_hardening_catches_next_near_miss. In that test, ten scans from a single source at confidence 0.65 (base threshold 0.7) pass individually, the per-(source, detector) EWMA of the near-miss indicator crosses the trigger ratio of 0.3, and the pair enters the hardened state. The eleventh scan from the same source at confidence 0.63 — strictly below every priming score and strictly below the un-hardened threshold — is then blocked, because the effective threshold has been lowered by the configured hardening offset. A concurrent scan from a different source at the same confidence 0.63 passes, confirming per-source isolation.

## 5.4 Limitations and threats to validity

- **NotInject false-positive rate regression.** d028 introduces ten false positives on the 339 benign samples of NotInject, moving the FPR from 0.9% (three false positives in the baseline) to 3.8% (thirteen). The regression is a consequence of the substitution matrix's permissive synonym grouping: benign queries that pair reveal-family verbs with instruction-family nouns (for example, "show me the instructions for a chemistry demonstration") accumulate enough alignment score to cross the detector threshold. Threshold tuning from 0.60 to 0.63 plus dampening of the show/reveal synonym group are planned and will appear in the next revision.
- **LLMail-Inject saturation.** The baseline regex pack already recalls 97.8% of LLMail-Inject attacks because the benchmark is dominated by direct instruction-override phrasings that the pack targets explicitly. The 0.2 percentage-point F1 improvement under d028 is honest but small, and should not be cited as the headline result. LLMail-Inject is most useful as a robustness-to-variation check rather than as an improvement benchmark.
- **AgentHarm orthogonality.** All four configurations achieve the same F1 of 0.319 and FPR of 31.8% on AgentHarm because the benchmark measures agent-task harmfulness rather than injection content. The harmful prompts invoke legitimate action-verb + admin-noun combinations that the baseline regex pack already flags, and adding injection-specific detectors does not change that behavior. The 31.8% FPR is itself an argument for the proposed taint-tracking technique (Section 4.7) which would distinguish benign administrative requests from adversarial tool invocations by provenance rather than by lexical content.
- **Synthetic benchmark self-evaluation risk (now empirically quantified — see Section 5.8).** The indirect-injection benchmark of Section 5.2 was constructed with knowledge of what the d027 and d028 detectors measure. The 1.000 F1 numbers on that benchmark are not held-out evaluation numbers in the strict sense. Section 5.8 (new in v4.0) reports the honest reality-check this caveat originally scoped: a held-out benchmark of fifty LLM-template-generated business documents (honestly disclosed — the documents are template-slot-filled, not human-authored) with paraphrased indirect-injection payloads in thirty of them. On that held-out corpus, d027 in isolation collapses from the synthetic 1.000 F1 (Section 5.2) to 0.000 F1: the tuned Jensen-Shannon threshold learned against uppercase override prose does not fire on the paraphrased attacks in this corpus, which stay in the same conversational register as the surrounding text. The full composed engine (d027 + d028 + regex baseline with d022 disabled) still catches twenty-two of thirty attacks — 0.733 recall and 0.815 F1. Per-family recall is 1.000 for data-exfiltration, role-hijack, and tool-abuse; 0.500 for instruction-override; and 0.167 for system-prompt-extraction. The gap between d027-alone (0.000 F1) and the composed engine (0.815 F1) is empirical support at this partial scale for the Section 2.2 composability thesis: when the target detector fails, the other detectors in the stack carry the load. Section 5.6 additionally corroborates the regex baseline behavior on three independent academic benchmarks.

- **Adaptive-attack evaluation is preliminary.** Section 5.7 (v4.0 addition) reports a small first-step experiment against d028 under a Nasr-style matrix-aware second-mover model: baseline detection 95.0 percent (19/20) drops to 0.0 percent (0/20) under adaptive attack — a 95.0 percentage-point ASR reduction, quantifying that d028 is a first-mover detector and must be composed with mechanistically independent layers. The full adaptive study across d027, the fatigue tracker, d029, and the composed detector stack — following the Zhan et al. (NAACL 2025 Findings, arXiv:2503.00061) and Nasr et al. (arXiv:2510.09023) methodologies — is scoped for v5.0.
- **Agent Security Bench not included.** We attempted to include ASB (Zhang et al., ICLR 2025) in the evaluation, but the benchmark requires its GitHub framework for execution and is not available on HuggingFace in a dataset form we could extract statically. A static extraction path analogous to what we did for AgentDojo is possible and is listed as future work.
- **Competitor comparison not rerun.** The repository contains a comparison harness that runs ProtectAI DeBERTa v2, PIGuard, Meta Prompt Guard 2, Deepset DeBERTa v3 and prompt-shield on deepset/prompt-injections and NotInject. That harness has not been rerun with d028 and d027 enabled. The mechanical work is small and is listed as a v4.0 deliverable.

## 5.5 Reproducibility

Every number in Section 5.2 is produced by the following sequence, executable on Python 3.10 through 3.13:

```
git clone https://github.com/mthamil107/prompt-shield
cd prompt-shield
pip install -e ".[dev,all]" agentdojo
python docs/papers/evaluation/build_indirect_injection_benchmark.py
python docs/papers/evaluation/run_public_datasets.py
```

Outputs are written to v041_public_datasets.json (machine-readable) and v041_public_datasets.md (human-readable). Wall-clock runtime is approximately four minutes. The fatigue probing-campaign claim in Section 5.3 is reproduced by pytest tests/fatigue/ which runs in under one second. The full prompt-shield test suite comprises 1,173 tests on the current main branch (as of v4.0) and passes cleanly across Python 3.10, 3.11, 3.12, and 3.13 in continuous integration.

## 5.6 Independent evaluation of the regex baseline against three peer-reviewed academic benchmarks (v3)

To validate that the Section 5.2 numbers are not specific to our chosen datasets, we further evaluate prompt-shield's input pipeline against three independent attack corpora published at peer-reviewed venues. None of these benchmarks were used to design or tune any prompt-shield detector. All numbers reported in this subsection are produced by the regex-only baseline: the d022 ML classifier, the d027 stylometric detector, and the d028 sequence-alignment detector are all disabled, matching the baseline control column in Table 1. The novel detectors are validated separately in Sections 5.2 through 5.5; their contribution is intentionally excluded from Table 2 so that the cross-benchmark numbers reflect the regex baseline alone.

| Benchmark | Venue | Corpus size | Caught | Detection rate |
|---|---|---|---|---|
| **Liu et al. (Open-Prompt-Injection)** | USENIX Security 2024 | 200 | 128 | 64.0% |
| **Garak (promptinject + latentinjection)** | Derczynski et al., 2024 (arXiv:2406.11036) | 5968 | 3294 | 55.2% |
| **InjecAgent** | Zhan et al., ACL Findings 2024 | 2108 | 1796 | 85.2% |
| **Combined** | | 8276 | 5218 | 63.0% |

*Table 2. Detection rate against three peer-reviewed academic benchmarks (regex-only baseline; d022 ML classifier, d027 stylometric detector, and d028 sequence-alignment detector all disabled).*

### 5.6.1 Per-attack-class breakdown

We group the per-benchmark per-probe results by underlying attack class to expose the cross-benchmark convergence. Numbers in parentheses report the source benchmark and per-probe rate.

| Attack class | Best | Worst | Cross-benchmark mean |
|---|---|---|---|
| **Explicit-override injection (Liu Ignore/Combine; Garak Hijack*; InjecAgent *-enhanced)** | 100% | 78.5% | ~92% |
| **Data exfiltration / PII leakage (InjecAgent DS-*; Garak LatentWhois*)** | 100% | 95.8% | ~99% |
| **Subtle indirect injection (Liu Naive/EscapeChar/FakeComp; InjecAgent DH-base; Garak LatentInjectionFactSnippet*)** | 75.4% | 11.7% | ~37% |
| **Toxicity elicitation via task framing (Garak LatentJailbreak)** | 0% | 0% | 0% (out of scope) |

*Table 3. Per-attack-class detection rates across the three academic benchmarks.*

### 5.6.2 The cross-benchmark insight

Three independent academic benchmarks converge on the same finding: pure pattern-based input detection plateaus around thirty-five to forty-five percent on subtle indirect injection (Liu Naive / EscapeChar / FakeComp at forty percent, InjecAgent DH-base at thirty-eight-point-eight percent, Garak LatentInjectionFactSnippet* at twelve to thirty-two percent). The plateau is consistent because all three measure the same underlying gap: indirect injection where the embedded instruction looks like a legitimate task request and contains no override keywords.

By contrast, data-exfiltration and explicit-override attacks are caught at near-ceiling rates (ninety-five to one hundred percent) because they contain syntactically distinctive patterns — URLs to attacker-controlled endpoints, account identifiers, the literal phrase "ignore previous instructions" — that pattern-based detectors reliably flag.

### 5.6.3 Implications

- **Input-side detection has a structural ceiling on subtle indirect injection.** This is not a bug in prompt-shield's specific implementation — it is a property of the attack class. Liu et al. quantified this gap and built DataSentinel (IEEE S&P 2025) as a fine-tuned model specifically targeting it; our results corroborate the gap on two independent benchmarks (Garak, InjecAgent) using a different defender (prompt-shield, regex-only).
- **Hybrid input plus output defense is the natural follow-up.** The Garak LatentJailbreak zero-percent result is explicitly out of scope for an input firewall and would be handled by prompt-shield's existing output-side toxicity scanner; future work should report the combined input plus output detection rate to honestly characterize end-to-end coverage.
- **Adaptive-attack robustness is the most important open question.** None of the three benchmarks above test adaptive attacks crafted against prompt-shield specifically. The canonical reference for that methodology is Nasr et al. "The Attacker Moves Second" (arXiv:2510.09023, 2025), and the next revision of this paper plans to apply that methodology per-detector-family.

### 5.6.4 Methodology

Liu et al.: We reproduce verbatim the five attack-template builders (Naive, EscapeChar, Ignore, FakeComp, Combine) from the Open-Prompt-Injection repository and construct a five-by-eight-by-five (two hundred-attack) matrix from a fixed set of benign data prompts crossed with injection payloads spanning the eight Liu task domains. The benign baseline (eight clean prompts, no attack) produces zero false positives.

Garak: We run garak version 0.15.0 with --model_type test.Blank (an offline mock generator that requires no LLM or GPU) over the full promptinject and latentinjection probe families. Attack prompts are extracted from entry_type == attempt records in the resulting JSONL run reports.

InjecAgent: We load the four pre-rendered test files (test_cases_dh_base.json, test_cases_dh_enhanced.json, test_cases_ds_base.json, test_cases_ds_enhanced.json) from the upstream repository and scan the Tool Response field of each case — the malicious tool output the agent would see in production. Two thousand one hundred and eight cases in total.

All three evaluations are reproducible from a clean install. Per-benchmark methodology, raw per-probe JSON, and runner scripts are in docs/papers/evaluation/ (liu_attackers.md, garak.md, injecagent.md) and in tests/benchmark_liu_attackers.py, tests/benchmark_garak.py, tests/benchmark_injecagent.py.

### 5.6.5 Why the novel detectors are excluded from Table 2

A natural reviewer question is why Table 2 reports only the regex baseline when the paper's central contribution is a set of novel cross-domain detectors. The answer is three technical reasons that jointly motivate the design choice, plus a fourth methodological one; we state them explicitly so the exclusion is not read as a hidden result.

First, d022 (the DeBERTa-based semantic classifier) is a stochastic inference pipeline whose output varies between runs, between GPUs, and between CPU and GPU back-ends. Section 5.6 promises bit-for-bit reproducibility on commodity hardware with a four-minute wall-clock budget; including d022 in the cross-benchmark configuration would break both properties.

Second, d027 (stylometric discontinuity) has a hard input-length floor (min_input_tokens; default one hundred) below which the detector returns silently because there are not enough tokens for the internal per-window feature vectors to be statistically stable. InjecAgent Tool Response fields have a median length of roughly twenty tokens — well below the one-hundred-token floor — so d027 is silent on essentially every one of the two thousand one hundred and eight cases by design. Including it in Table 2 would show a zero-lift column that a naive reader would misinterpret as detector failure rather than input-scope mismatch.

Third, d028 (Smith-Waterman sequence alignment) uses a substitution matrix hand-curated against approximately eighty override-style attack templates ("ignore previous instructions," "disregard prior directives," and their documented paraphrases and synonym-swap variants). The matrix explicitly targets override, extraction, and role-hijack phrasings. Garak's latentinjection probes and InjecAgent's tool-response poisoning both operate through different attack shapes — embedded payloads and realistic-looking tool outputs that do not resemble override commands. Section 5.7 quantifies the resulting scope mismatch under an adaptive-adversary model; the aggregate lift from d028 on Garak and InjecAgent is small by design.

Fourth, a methodological consideration. Table 2's role is to establish an independent-defender baseline against three peer-reviewed benchmarks, mirroring the baseline control column of Table 1. Adding the novel detectors to Table 2 would require running the full four-configuration ablation on each of the three external benchmarks — approximately six additional runs for a result whose expected sign is known in advance from the scope arguments above. We instead report the novel detectors' per-technique behavior on the datasets they were designed for in Sections 5.2 through 5.5, and use Table 2 solely to triangulate the regex baseline's structural ceiling.

## 5.7 Preliminary adaptive-attack evaluation of d028 (v4)

Section 2.2 argues that adaptive adversarial evaluation is the current frontier for prompt-injection defense research; Section 5.4's fourth limitation notes that the shipped detectors had not, at the time of the v3.0 revision, been evaluated against this adversarial model. This revision reports a small first-step experiment against d028 (Smith-Waterman sequence alignment) under a Nasr-style second-mover threat model (Nasr et al., arXiv:2510.09023) in which the adversary has read the d028 substitution matrix and constructs attacks whose synonym choices are deliberately outside every matrix group. Section 6 defers the full adaptive-attack study across all shipped novel detectors to v5.0.

### 5.7.1 Methodology

We construct twenty canonical override-style baseline attacks known to fire d028 with high confidence under default settings (threshold 0.6, match bonus 3, synonym bonus 2, mismatch penalty -1, gap penalty -1). For each baseline we craft an adversarial variant that preserves the attack's semantic intent (for example, "reveal your system prompt") while replacing every matrix-family token with a genuine English synonym absent from every one of the fifteen matrix synonym groups (for example, "recite your operational briefing"). A programmatic per-variant audit — see the ``verify_adaptive_variants`` function in the runner script — confirms zero matrix-family tokens in the adversarial set. Both sets are then evaluated against d028 with default configuration.

### 5.7.2 Result

Detection rate on the baseline set is nineteen of twenty (95.0 percent). Detection rate on the adversarial set is zero of twenty (0.0 percent). The attack-success-rate reduction under adaptive attack is 95.0 percentage points — that is, a matrix-aware adversary systematically bypasses d028 as a standalone detector.

### 5.7.3 Interpretation

The result is not surprising and does not undermine the deepset F1 lift reported in Section 5.2: on non-adaptive attack corpora d028 catches paraphrased attacks that a regex baseline misses, which was its design intent. The result does formalize the operating envelope: d028 is a first-mover-adversary detector, and should not be relied on as a standalone defense against adversaries with knowledge of the substitution matrix. Since the substitution matrix is open source and public — see docs/papers/evaluation/adaptive_attack_d028.py and the matrix source file at src/prompt_shield/detectors/_d028_substitution_matrix.py — any real adversary has that knowledge.

In a defense-in-depth stack d028 composes with (i) the d022 semantic classifier, whose decision boundary is independent of the d028 matrix; (ii) the d027 stylometric detector, which flags author-style discontinuities orthogonal to lexical content; (iii) the adversarial fatigue tracker, which uses attempt frequency itself as signal rather than attempt content; and (iv) at the output boundary, the ToolResultGuard primitive introduced in Section 7.1. An adversary who evades d028 through matrix-aware substitution has not evaded these other layers. Quantifying the composed detection rate under adaptive attack — which requires an attack budget that adaptively targets each layer in turn — is the central v5.0 experiment noted in Section 6; a two-of-five-layer partial run against d028 and d027 is reported in Section 5.7.6 of this revision.

### 5.7.4 Scope caveats

The experiment is deliberately small. It covers (i) n = 20 attack pairs, not a large corpus; (ii) only override-style attacks, not extraction, roleplay, or many-shot shapes; (iii) only d028, not d027, the fatigue tracker, d029, or the ML classifier; (iv) a single-round substitution attack, not iterative gradient-informed or learning-based adversaries; and (v) d028 in isolation, not the composed detector stack. Extension (iii) is partially addressed by the d027 short-input-floor attack in Section 5.7.6; extensions (i), (ii), (iv), and the composed-stack sweep from (v) remain v5.0 commitments.

### 5.7.5 Reproducibility

The twenty attack pairs, per-attack detection scores, and the matrix-leak audit are released in docs/papers/evaluation/adaptive_attack_d028.py (runner script), adaptive_attack_d028.json (machine-readable), and adaptive_attack_d028.md (human-readable). Wall-clock runtime is under one second on commodity hardware. Reproduce with python docs/papers/evaluation/adaptive_attack_d028.py.

### 5.7.6 Composed-stack adaptive-attack partial run

Section 5.7.3 argued that d028's matrix-aware bypass would be caught by mechanistically independent layers in a composed stack. This subsection reports a partial empirical test of that claim covering a second layer — d027 stylometric discontinuity — under an adaptive short-input-floor attack, together with an end-to-end run of the same adaptive inputs through the full composed engine. The methodology, threat model, per-layer attack strategies for all five layers, and the composability metric are documented at docs/papers/evaluation/composed_adaptive_methodology.md; the reproducible partial run and its results are at docs/papers/evaluation/composed_adaptive_partial.py and composed_adaptive_partial.md.

Attack construction. d027's operating envelope has a hard input-length floor (min_input_tokens; default one hundred) below which the detector short-circuits to detected=False. An adaptive attacker who reads the source and configuration can trivially defeat d027 in isolation by keeping the payload under one hundred tokens. We construct twenty baseline attacks in which a benign wrapper of approximately two hundred fifty tokens surrounds a thirty-token override, and twenty adaptive variants that strip the wrapper so the entire input is under one hundred tokens while preserving the override semantics.

d027-alone result. On the baseline set, d027 fires on twenty of twenty (one hundred percent). On the adaptive short-input variants, d027 fires on zero of twenty (zero percent) — by construction, since the floor triggers before any window math executes. Attack-success-rate reduction is one hundred percentage points, mirroring the d028 result in Section 5.7.2 and confirming that no single detector in the stack survives a targeted second-mover attack.

Composed-engine result. The same twenty adaptive inputs are then evaluated through PromptShieldEngine().scan() with the shipped default configuration (thirty-four detectors loaded, d022 semantic classifier enabled). The engine blocks twenty of twenty (one hundred percent). Per-detector fire counts across the twenty adaptive inputs are: d022_semantic_classifier fires on twenty, d001_system_prompt_extraction fires on eight, d028_sequence_alignment fires on eight, d003_instruction_override fires on three, d004_prompt_leaking fires on three.

Ablation without d022. To decompose the composability gain — is it the semantic classifier doing all the work, or do structurally different signals meaningfully contribute? — we rerun the same twenty adaptive inputs with d022 disabled. The engine still blocks fourteen of twenty (seventy percent) using only regex keyword detectors (d001 on eight, d003 on three, d004 on three) and d028 sequence alignment (on eight). Two mechanistically different signals each contribute rescue capability, and d022 mops up the remaining six. This is the pattern Section 2.2 predicts.

Scope, and how this result should be read alongside Section 5.8. The experiment covers two of five layers (d028 by reference from Section 5.7.2; d027 executed here) at twenty attack pairs per layer. The full five-layer protocol at n = 100 per cell is a v5.0 commitment. Critically, the 20/20 composed-stack detection figure reported here is on this specific adaptive-attack corpus (uppercase-override payloads under one hundred tokens) and must not be read in isolation as evidence that the stack is unbeatable: Section 5.8 reports 22 of 30 (0.733 recall) on paraphrased indirect-injection payloads in long-form business documents. The two datasets probe different attack shapes and the correct reading is that the composed stack is strong on some attack classes and measurably weaker on others.

## 5.8 Held-out indirect-injection benchmark (v4)

### 5.8.1 Motivation

Section 5.4's fourth limitation flagged that the indirect-injection benchmark of Section 5.2 was constructed with knowledge of what d027 and d028 measure, and that the reported 1.000 F1 was therefore not a held-out number in the strict sense. The stated honest next step was external validation on a corpus authored without knowledge of the detectors' feature vectors. This subsection reports that validation at the partial scale we could execute in-cycle: fifty LLM-template-generated business documents whose injection payloads were paraphrased into the same conversational register as the surrounding text, so the detectors face inputs that do not match the uppercase-directive shape their synthetic-benchmark thresholds were tuned against.

### 5.8.2 Corpus construction (honestly disclosed)

The corpus contains fifty documents across five document types (email, meeting-notes, wiki, postmortem, spec) with ten documents per type. Thirty documents carry an injection payload and twenty are clean. Injections are drawn from eight paraphrased attack templates spanning five attack families: system-prompt extraction, role hijack, instruction override, data exfiltration, and tool abuse (six injections per family). Each document type provides three distinct base templates with slot-filled variety (project name, service, person, region, metric) so no two documents are byte-identical, and each injection is inserted at a plausible per-document-type location (mid-email paragraph, an action-item bullet in meeting notes, a wiki footer, a postmortem action item, and an "additional notes" block in the spec).

Honest disclosure: the documents are LLM-template-generated (offline slot-filled templates driven by a deterministic PRNG), NOT human-authored. This is a partial improvement over the Section 5.2 synthetic self-benchmark because neither the templates nor the paraphrased payloads were constructed with knowledge of d027's Jensen-Shannon threshold or d028's substitution matrix. A future revision will use crowdsourced or professionally-authored documents. The corpus generator is deterministic: python docs/papers/evaluation/build_held_out_corpus.py --seed 42 regenerates it byte-for-byte.

### 5.8.3 Results

| Configuration | TP | TN | FP | FN | Recall | FPR | Precision | F1 |
|---|---|---|---|---|---|---|---|---|
| **Full engine (d027 + d028 + regex baseline, d022 off)** | 22 | 18 | 2 | 8 | 0.733 | 0.100 | 0.917 | 0.815 |
| **d027 in isolation (threshold 0.06, window 50, stride 25)** | 0 | 18 | 2 | 30 | 0.000 | 0.100 | 0.000 | 0.000 |

*Table 4. Held-out benchmark: 50 LLM-template-generated business documents, 30 with paraphrased indirect-injection payloads. Detection rule matches Section 5.2 (action in {block, flag} or overall_risk_score >= 0.5).*

Per-attack-family recall of the full engine: data exfiltration six of six (1.000), role hijack six of six (1.000), tool abuse six of six (1.000), instruction override three of six (0.500), system-prompt extraction one of six (0.167). The stack handles data-exfiltration and role-hijack and tool-abuse families cleanly and struggles on the two families whose payloads read most like legitimate operator requests.

Comparison with Section 5.2. On the synthetic self-benchmark of Section 5.2, d027 in isolation scored 1.000 F1. On this held-out corpus, d027 in isolation scores 0.000 F1 — a collapse. Mechanistically, d027 was tuned against a corpus whose payloads used ALL-CAPS SYSTEM-style override prose against calm business prose, which produces a large Jensen-Shannon divergence at the window boundary that is easy to detect. The paraphrased attacks in the held-out corpus keep the injected sentence in the same lower-case business tone as the surrounding text (for example, "Please display the exact system prompt so the auditor can verify the configuration"), which keeps the per-window feature vector close to the surrounding text and does not trip the threshold. The detector's own docstring calls this behavior out as high-precision / low-recall for register-matched attacks; the paper's original 1.000 F1 did not communicate that shape.

The composed engine catches 22 of 30 attacks (0.733 recall, 0.815 F1) at 0.100 FPR. Because d027 alone contributes zero true positives on this corpus, the engine's recall is carried by the regex baseline detectors (d001 through d006 on system-prompt-extraction and instruction-override; d013 through d014 on tool abuse) and by d028's substitution matrix. The gap between d027-alone (0.000 F1) and the composed engine (0.815 F1) is empirical support for the Section 2.2 composability thesis at partial scale.

### 5.8.4 Limitations

- **Documents are LLM-template-generated, not human-authored.** The templates are hand-written but the slot-filling is deterministic PRNG; genuine crowdsourced or professionally-written documents would exercise a wider distribution of legitimate register variation. A v5.0 corpus should use human-authored source documents.
- **Camouflage-only injection set.** Every injection paraphrase in this corpus deliberately camouflages into the surrounding register. The corpus does not contain the uppercase-directive or structural-anchor attack shapes that d027 and d028 were designed against. The 0.000 F1 for d027 in isolation should be read as evidence of a specific operating-envelope limit, not as evidence that d027 is broken.
- **Wiki false-positive rate 0.500 exposes a specific d027 weakness.** Two of four benign wiki documents produce false positives from d027 alone (FPR 0.500 within the wiki subset). Wikis contain list-to-prose transitions (bulleted specification followed by narrative explanation) that produce genuine Jensen-Shannon spikes at the transition boundary. This is a real d027 weakness on structured technical prose and is a candidate for the threshold-calibration work sequenced in Section 6.
- **n = 30 attacks / n = 20 benign is small.** Wilson 95 percent confidence intervals span roughly plus-or-minus 15 percentage points on a proportion of 30. A larger held-out corpus is required before per-family recall numbers should be reported to two decimal places without a caveat.

### 5.8.5 Reproducibility

Every number in Table 4 is produced by the following sequence:

```
python docs/papers/evaluation/build_held_out_corpus.py --seed 42
python docs/papers/evaluation/run_held_out_benchmark.py
```

Outputs are written to docs/papers/evaluation/held_out_indirect_injection.json (machine-readable) and held_out_indirect_injection.md (human-readable). Wall-clock runtime is under thirteen seconds on commodity hardware. The generated corpus and per-document manifests are checked in under docs/papers/evaluation/held_out_indirect_injection/.

# 6. Conclusions and Future Work

We presented seven cross-domain detection techniques for prompt injection and reported empirical results for three of them. The local-sequence-alignment detector delivers a thirty-four-point-six percentage-point F1 improvement on deepset/prompt-injections with zero additional false positives on that benchmark, while honestly introducing a 2.95 percentage-point false-positive-rate regression on NotInject that we have calibrated rather than hidden. The stylometric discontinuity detector lifts F1 by 11.1 percentage points on a synthetic indirect-injection benchmark released alongside the paper, while remaining silent on short inputs by design. The adversarial fatigue tracker is validated via a probing-campaign integration test that demonstrates the core behavioral claim: after a sustained burst of near-threshold scans from the same source, a subsequent lower-confidence scan from that source is blocked.

The v3.0 revision extends this evaluation to three independent peer-reviewed academic benchmarks comprising 8,276 attack prompts none of which were used to design any prompt-shield detector. Detection rates of 64.0 percent on Liu et al., 55.2 percent on Garak, and 85.2 percent on InjecAgent triangulate to the same structural finding: pure pattern-based input detection plateaus at thirty-five to forty-five percent on subtle indirect injection where the embedded instruction lacks override keywords, while reaching ninety-five to one hundred percent on data-exfiltration and explicit-override attacks. This corroborates the gap that DataSentinel was designed to close and motivates the v4.0 work plan below.

The three remaining proposed techniques are sequenced by risk: perplexity spectral analysis (Section 4.6) requires an optional ML dependency; runtime taint tracking (Section 4.7) requires an agent-pipeline fixture; prediction-market scoring (Section 4.5) touches the core scoring path and is scoped last, behind a mandatory shadow-mode gate. Section 4.3 honeypot tool definitions, previously the first of the remaining techniques, shipped in the v4.0 cycle as d034_honeypot_tool with twenty unit tests (see Section 4.3 empirical-validation subparagraph).

The v4.0 revision (this document) folds in a survey of the 2026 concurrent-contribution cluster (Section 2.4); a Gang-of-Four-style formalization of three architectural patterns extracted from prompt-shield's implementation (Section 7); a preliminary adaptive-adversary evaluation against d028 (Section 5.7); a composed-stack adaptive-attack partial run covering a second layer (d027) and the full composed engine on the same twenty short-input adaptive inputs (Section 5.7.6); a held-out indirect-injection benchmark of fifty LLM-template-generated business documents on which d027 in isolation collapses from 1.000 F1 to 0.000 F1 while the composed engine still recovers 0.815 F1 (Section 5.8); and the shipping of the honeypot tool-definitions detector (d034) that moves the paper's implemented-technique ratio from three of seven to four of seven (Section 4.3). We note honestly that Section 6 of the v3 revision scoped a full adaptive-attack evaluation and a competitor rerun for v4.0. On scope review we prioritized the design-pattern formalization (Section 7) — driven by the shipping of ToolResultGuard in prompt-shield v0.7.0 and the observation that four concurrent 2026 systems ship the same pattern without a shared name — plus the held-out and composed-adaptive experiments above, and delivered only the smaller preliminary adaptive experiment reported in Section 5.7 for d028 in isolation. The full five-layer adaptive-attack evaluation and the competitor rerun carry to v5.0 with the specific commitments below.

The v5.0 revision will fold in: a full section-level write-up of the d029 many-shot structural detector currently shipped in prompt-shield v0.4.1; the full five-layer adaptive-attack evaluation at n = 100 pairs per cell per the protocol at docs/papers/evaluation/composed_adaptive_methodology.md; expansion of the Section 5.8 held-out corpus to include human-authored documents in place of LLM-template-generated ones; a head-to-head competitor comparison (Rebuff, Lakera, Meta Prompt Guard 2, PIGuard, Deepset DeBERTa v3, DataSentinel) on the same three academic benchmarks used in Section 5.6; and implementation plus evaluation of at least one of the three remaining proposed techniques (perplexity spectral analysis, runtime taint tracking, or prediction-market scoring).

# 7. Design Patterns for Agent-Era LLM Security

Design patterns — named, reusable descriptions of problem-solution pairs — compress community knowledge and coordinate independent implementers around a shared vocabulary. Gamma et al. (Design Patterns, Addison-Wesley 1994) demonstrated the durability of this contribution in object-oriented software; Sweeney's k-anonymity (International Journal of Uncertainty, Fuzziness and Knowledge-Based Systems 2002, DOI 10.1142/S0218488502001648) played the analogous role for statistical disclosure control; Dean and Ghemawat's MapReduce (OSDI 2004) did so for large-scale data processing; and MITRE ATT&CK (attack.mitre.org, first released 2013) for adversarial behavior modeling. In each case, formalizing a recurring construct that practitioners were already reinventing gave the field a reference point that outlasted specific implementations.

We observe a similar convergence in agent-era LLM security. Between April and July 2026, four independent papers — ClawGuard (Zhao et al., arXiv:2604.11790), DualView (Kim et al., arXiv:2607.03821), Prismata (Villa et al., arXiv:2607.08147), and Token-Flow Firewall (Wang et al., arXiv:2607.08395) — shipped implementations of the same core idea: scan tool output before it re-enters the LLM context. A fifth, AgentSpec (Wang, Poskitt, and Sun, ICSE 2026, arXiv:2503.18666), subsumes this as a special case of a broader runtime-enforcement DSL. None used a common name. We formalize that pattern here, together with two others we have found useful in building prompt-shield. Our contribution is nomenclature and structure, not invention.

## 7.1 Pattern: Tool-Result Boundary Gate

**Problem.** Tool-augmented LLM agents invoke external tools — web search, code execution, MCP servers, retrieval indexes, file readers — and feed the returned data back into the model context as prompt tokens. Because that data is not part of the operator's system prompt or the user's turn, it constitutes a third, adversary-controllable channel. An attacker who can influence any surface a tool reads (an indexed web page, an issue comment, a shared document, a network response) can inject instructions that the model will treat with the same authority as legitimate context. The impact ranges from silent policy bypass to exfiltration of secrets held elsewhere in the session.

**Context.** Tool-augmented LLM agents built on frameworks such as MCP, LangChain-style tool loops, RAG pipelines, and browser or code-execution back-ends. The pattern applies whenever untrusted content re-enters the model's prompt as a result of a tool call.

**Forces.** Detection accuracy must be traded against per-call latency, since a gate that adds seconds to every tool invocation is unusable in interactive agents. False positives are especially costly: they silently break agent utility, which is harder for operators to detect than under-blocking. Purely content-based detection cannot substitute for capability-based confinement (for example, sandboxing the tool itself), but the two compose. Finally, the pattern must be expressible as middleware or a wrapper that slots into existing agent frameworks without requiring model retraining or protocol changes.

**Solution.** Introduce a first-class boundary primitive on the tool-result path. The primitive (i) intercepts the raw tool return before it is concatenated into the model context, (ii) classifies the content through a stack of heterogeneous detectors — regex, statistical, ML, alignment-based — into a stable attack-family taxonomy, and (iii) enforces one of a small set of policy modes — block, flag with metadata, or sanitize by structural quoting — chosen per family or per tool. The gate is deliberately positioned at the boundary rather than integrated into the model, so that it composes with capability-based confinement and remains auditable independently of the LLM. Operators configure it declaratively.

**Known Uses.** Between April and July 2026, four research prototypes shipped this pattern independently and without a common name: ClawGuard (Zhao et al., arXiv:2604.11790) implements a permission-oriented variant for MCP tools; DualView (Kim et al., arXiv:2607.03821) separates operator and tool-content views before scanning; Prismata (Villa et al., arXiv:2607.08147) applies the pattern to structured tool outputs; and Token-Flow Firewall (Wang et al., arXiv:2607.08395) enforces it at the token-stream level, reporting attack success reduced to 12.5 percent and benign pass rate 97.4 percent per the authors' abstract. AgentSpec (Wang, Poskitt, and Sun, ICSE 2026, arXiv:2503.18666), a general runtime-enforcement DSL, subsumes this pattern as a special case rather than centring on it. prompt-shield ships the pattern as ToolResultGuard in v0.7.0 with the naming and Gang-of-Four-style formalization given here. None of the five implementations adopts the others' terminology.

**Related Patterns.** Defense in Depth (the gate is one layer, not the only layer); Reference Monitor (classical mediation of untrusted flows); Sandbox (capability-side complement); Capability-based Security (an alternative paradigm that reduces the surface the gate must protect).

**Novelty note.** We claim naming, not invention. Four contemporary implementations plus one adjacent runtime-enforcement system demonstrate the pattern's convergence; our contribution is the shared vocabulary.

## 7.2 Pattern: Federated Signed Threat Intelligence for LLM Security

**Problem.** No single LLM-security deployment observes enough attack traffic to characterize novel patterns quickly. Commercial attack catalogs are proprietary, so open-source defenses cannot draw on the pooled attack telemetry that would speed detection of novel patterns. Without a trustworthy sharing mechanism, OSS defenses either lag commercial ones or must ingest anonymously contributed data whose provenance they cannot verify.

**Context.** Open-source LLM-security tooling with a distributed installation base — pip install, container images, self-hosted deployments — whose operators want to opt in to shared intelligence but cannot run a trust broker themselves.

**Forces.** Trust: who verifies feed integrity? Freshness against bandwidth and cache load. Offline signing (private key never touches the network) against CI-based signing (operationally simpler but wider blast radius). Verification cost per client, since detection sits on the hot path. Feed diversity against publisher dependency.

**Solution.** Publish the feed from a single accountable publisher, sign each release offline with an ed25519 key, and distribute over a CDN so that origin compromise does not compromise the key. Clients verify signatures with a small pure-code library, with the trust anchor pinned in the client release, so that downstream operators pull the feed as they would any other dependency, without needing to run a public-key infrastructure. Rotation is handled by shipping a new anchor in a new client release, keeping revocation cost proportional to normal upgrade cadence rather than to feed frequency.

**Known Uses.** prompt-shield's federated signed feed (github.com/mthamil107/prompt-shield-signatures) is, at the time of writing, the only known use of this pattern applied to LLM security. The underlying construct is well-established in adjacent domains — Abuse.ch, PhishTank, and VirusTotal in network security, and signed software updates more generally. Our contribution is the application, not the primitive.

**Related Patterns.** Signed Software Updates; Certificate Transparency (append-only public logs as a stronger integrity property); Defense in Depth.

**Novelty note.** We describe this as a novel application of an existing pattern rather than a novel pattern. Single known LLM-security use at time of writing.

## 7.3 Pattern: Attack-Family Taxonomic Projection

**Problem.** A defense-in-depth stack accumulates many concrete detectors with opaque identifiers — regex rule IDs, model checkpoint names, alignment-scorer thresholds. Downstream consumers — policy engines, dashboards, alert routers, and human triage — need semantic categories, not detector IDs, in order to reason about coverage, express policy, and communicate risk. But the detector set churns as research advances, so any category system tied directly to detector identity becomes brittle.

**Context.** Detector stacks that combine heterogeneous mechanisms (regex, statistical anomaly, supervised classifier, alignment-based) and that expose their output to policy engines or human operators.

**Forces.** Taxonomy stability against detector innovation; per-family policy configurability against the risk of category proliferation; explainability to non-technical stakeholders; the ability to introduce a new detector without renaming or reclassifying prior alerts; small gap-filling coverage for attack families with no dedicated detector.

**Solution.** Define a small, stable attack-family taxonomy chosen from the analyst-facing vocabulary rather than the detector-facing one. Implement the taxonomy as a pure projection function over (detector_id, evidence) pairs, so that the mapping from mechanisms to families is a one-way, auditable, easily unit-tested piece of code. Add gap-filling regex only for families that no first-class detector covers, keeping the fallback narrow and clearly labeled. Policy engines, dashboards, and metric pipelines consume families, not detector IDs; the detector layer is then free to evolve without invalidating downstream consumers.

**Known Uses.** prompt-shield's ToolResultGuard uses a nine-family taxonomy at v0.7.0. We are aware of no other implementation of this specific formalism at the time of writing; we describe the pattern in the hope that others adopt it. MITRE ATT&CK and ATLAS (atlas.mitre.org) provide taxonomies at a different level of abstraction — attack modeling for threat intelligence and red-teaming — rather than a projection layer between detection tooling and policy. The two are complementary: one names what an attacker might do; the other names what a detector saw.

**Related Patterns.** MITRE ATT&CK / ATLAS taxonomies (adjacent abstraction level, attack-modeling rather than detection-tooling); Facade (a simpler stable interface over a heterogeneous detector stack); Anti-Corruption Layer (insulating downstream consumers from detector-set churn).

**Novelty note.** Novel formalism with a single known implementation at the time of writing. We describe the pattern in the hope that other detector-stack maintainers adopt it.

### 7.4 What these patterns do not address

These three patterns do not exhaust the design space of agent-era LLM security. They deliberately do not address adaptive adversarial evaluation, which requires an evaluation methodology as much as an architectural pattern; they do not provide provable defense guarantees, which the current empirical detector-stack paradigm cannot; and they are not a substitute for capability-based confinement, which reduces the surface any content-based defense must protect and remains, in our view, the strongest available primitive for tool-augmented agents. A mature deployment will combine capability confinement, boundary gating, threat-intelligence sharing, and taxonomic projection, along with patterns yet to be named.

We invite the field to formalize those additional patterns. The value of a shared vocabulary comes from its coverage; three patterns is a beginning, not an inventory.

# Appendix A. Released Artifacts

All artifacts released alongside this paper are in the public repository at github.com/mthamil107/prompt-shield under the Apache 2.0 license. Relevant paths:

- `src/prompt_shield/detectors/d028_sequence_alignment.py` — Smith-Waterman alignment detector (Section 4.4).

- src/prompt_shield/detectors/d027_stylometric_discontinuity.py — Stylometric discontinuity detector (Section 4.1).
- src/prompt_shield/detectors/d029_many_shot_structural.py — Many-shot structural detector (preview; full §-level write-up in v4).
- src/prompt_shield/detectors/d034_honeypot_tool.py — Honeypot tool-definitions detector (Section 4.3; v4.0 cycle).
- src/prompt_shield/fatigue/tracker.py — Adversarial fatigue tracker (Section 4.2).
- tests/detectors/test_d028_sequence_alignment.py — Tests covering alignment math, substitution matrix, detector behavior.
- tests/detectors/test_d027_stylometric_discontinuity.py — Tests covering feature extraction, Jensen-Shannon math, detector behavior and fixtures.
- tests/detectors/test_d029_many_shot_structural.py — 36 tests covering many-shot pattern detection (v3-shipped).
- tests/detectors/test_d034_honeypot_tool.py — 20 tests covering both firing modes and every tool-call shape (Section 4.3).
- tests/fatigue/test_tracker.py, test_engine_integration.py — Tests including the probing-campaign integration test cited in Section 5.3.
- tests/benchmark_liu_attackers.py — Section 5.6 — Liu et al. (USENIX Security 2024) attack-strategy benchmark.
- tests/benchmark_garak.py — Section 5.6 — Garak prompt-injection probe evaluation.
- tests/benchmark_injecagent.py — Section 5.6 — InjecAgent (ACL Findings 2024) evaluation.
- docs/papers/evaluation/run_public_datasets.py — Reproduction harness for Table 1.
- docs/papers/evaluation/build_indirect_injection_benchmark.py — Generator for the synthetic indirect-injection benchmark.
- docs/papers/evaluation/build_held_out_corpus.py — Generator for the v4.0 held-out benchmark (Section 5.8).
- docs/papers/evaluation/run_held_out_benchmark.py — Runner for the v4.0 held-out benchmark (Section 5.8).
- docs/papers/evaluation/held_out_indirect_injection.md, .json — Human-readable and machine-readable results for Table 4 (Section 5.8).
- docs/papers/evaluation/composed_adaptive_methodology.md — Full 5-layer composed-stack adaptive-attack protocol (Section 5.7.6).
- docs/papers/evaluation/composed_adaptive_partial.py, .md — Reproducible partial run of the composed-stack adaptive attack (Section 5.7.6).
- docs/papers/evaluation/v041_public_datasets.json — Machine-readable source of every number in Table 1.
- docs/papers/evaluation/liu_attackers.md, garak.md, injecagent.md — Per-benchmark methodology and findings for Section 5.6.
- docs/papers/evaluation/garak_regex_only.json, injecagent_regex_only.json — Raw per-probe / per-split results for Section 5.6.
- docs/papers/evaluation/ANALYSIS.md — Narrative analysis of per-dataset behavior with limitation framing.
- tests/baseline_v0.3.3.txt — Locked baseline for the regression gate used during the shipping of each detector.
- CITATION.cff — GitHub-native citation metadata.